# Predicting the electroporated tissue area trajectory in Electroporation-based protocol optimization

Guillermo Marshall[1,*], Alejandro Soba[2]

[1] Laboratorio de Sistemas Complejos, Departamento de Computación, Instituto de Física del Plasma, Departamento de Física, Facultad de Ciencias Exactas y Naturales, Universidad de Buenos Aires, Consejo Nacional de Investigaciones Científicas y Técnicas.
[2] Comisión Nacional de Energía Atómica, Buenos Aires, Argentina.

## Abstract

Electroporation (EP), the temporary or permanent permeabilization of the cell membrane induced by an electric field, is the basis of a wide range of applications in medicine and food processing. In EP-based protocol optimization in terms of pulse number, such as in electrochemotherapy (ECT), irreversible electroporation (IRE), and gene electrotransfer (GET), it is essential to reach an optimal dose-response, that is, the pulse dose that maximizes the electroporated tissue area while minimizing damage. The prediction of the electroporated tissue area variation in time, i.e., its trajectory, requires measuring the EP threshold trajectory as well as understanding the interaction between both trajectories and the damaged tissue area trajectory. Here we introduce a new methodology, based on the analysis of the nonlinear dynamic interaction of the EP threshold and damaged tissue area trajectories, that shows that the EP threshold trajectory is the time gradient of the electric field. This allows predicting the electroporated area trajectory with a single electroporated area measurement at the last pulse, thus avoiding the need to measure the EP threshold trajectory, a rather cumbersome task. Also, it makes it possible to explain at the macroscopic level why the EP threshold trajectory has an approximate exponential time decrease while the EP threshold isoline trajectory, aka the electroporated tissue area trajectory, has an approximate logarithmic time increase. Further, it permits predicting an optimal dose response in terms of pulse number in an EP-based protocol, with a single electroporated tissue area measurement. Examples of its application to an in vitro vegetal model, and to an in vivo skinfold chamber shed new light on the nonlinear dynamic behavior of electroporated tissue area, EP threshold, and damaged tissue area trajectories, paving the way for optimal treatment planning in EP-based protocols.

\* Corresponding author
  E-mail: gmarshall@dc.uba.ar

# Introduction

Electroporation (EP), the temporary (reversible), or permanent (irreversible) permeabilization of the cell membrane induced by the application of an electric field, is the basis of a broad range of applications in medicine and food processing, [1]. In medicine, the EP-based tumor treatments that stand out among others are electrochemotherapy (ECT), irreversible electroporation (IRE), and gene electrotransfer (GET).

Numerous attempts have been made since the inception of electroporation [2] for optimizing EP-based protocols in terms of electrical parameters; various parameters are considered: pulse duration, frequency, number of pulses, applied voltage, number of electrodes, and their placement, among others. In this context, mathematical and computational modeling validated with experimental data became a powerful tool. The *Standard EP model,* i.e., the formation of aqueous pores in the cell membrane when the electric field reaches a threshold, is the cornerstone of electroporation. According to the *Standard EP model,* predicting the outcome of an EP-based protocol in terms of electrical variables consists of finding the extent of the electroporated tissue area that is, computing the electric field, and matching it with a known threshold. In mathematical terms, this translates into solving the nonlinear Poisson equation for the electric field distribution, and its matching with the threshold from experimental measurements. Its use is generalized for obtaining the threshold value at the last pulse if the electroporated area at the last pulse is known.

In a pioneering in vivo GET protocol in [3] an optimal GET plasmid DNA transfer in muscle fibers was reported: eight pulses of 20 ms at 1 Hz with a voltage-to-distance ratio of 200 V/cm. [4] analyzed the influence of pulse parameters on GET protocols, using an identical setup as in [3]. Results showed that the uptake increased with the electric field up to a range of 300-400 V/cm and then decreased, the uptake increased with pulse length in the range of 5-25 ms, and with pulse number in the range of 4-32 pulses. It was concluded that the EP threshold for eight pulses of 20 ms at 1 Hz, was around 100 V/cm. The EP threshold was obtained with the *standard EP model* [5] using plate electrodes. In summary, the authors suggested the use of eight pulses of 20 ms for GET protocols in tissue, obtaining the EP threshold under those conditions.

ECT and GET protocols (8 pulses of 100 µs at 1Hz, and various intensities in the range of 860-1360 V) validated with an in vivo model were presented in [6]. The authors made the first precise determination of in vivo reversible and irreversible tissue EP thresholds with the *standard EP model*: 362 V/cm, and 637 V/cm, respectively. Following previous work, [7] presented ECT and GET protocols simulations with the *standard EP model* (8 pulses of 100 µs at 1Hz, and various intensities in the range of 200-1200 V) validated with an in vivo model. The authors reported reversible and irreversible tissue EP thresholds of 460 V/cm and 700 V/cm, respectively. An extensive study review of in vivo, in vitro, in silico, and translational IRE treatment protocols was presented by [8]; authors reported that most in vivo studies found the EP threshold to be in the range of 500–1000 V/cm for electric field intensities in the range of 1000-2000 V/cm. The EP thresholds were determined with the *standard EP model*.

Based on the determination of the relation between the amplitude and duration of pulses resulting in the same electroporated tissue area, [9] introduced the equivalent pulse parameter concept. The authors found that with a higher number of pulses, lower EP thresholds were necessary to obtain identical electroporated tissue areas. An in vitro tumor platform for modeling IRE was analyzed by [10]; for voltages in the range of 150-600 V/cm, the maximum conductivity obtained was in the range of 1.2-1.9 S/m with a baseline of 1.2 S/m. The increase in conductivity had almost no effect on the electric field distribution, and thermal changes had minimal effect on the electric field distribution;

the IRE EP threshold of 500 V/cm for cell death was determined with the *standard EP model*. [11] studied the optimization of the electric field distribution for different electrical parameters and electrodes using a genetic algorithm. It was shown that parallel array electrodes are best suited for spherical tumor geometry because the whole tumor is subjected to a sufficiently high electric field while requiring minimum electric current and causing minimum tissue damage. [12] studied the influence of pulse number in an EP-based vegetable model with experiments and simulations using the *standard EP model* (the electrical conductivity was a function of the pulse number, as well). The authors showed that the electric current and the electric conductivity increased proportionally with the pulse number, while the electric field remained constant (after the first pulse). This is because the pulse intensity is kept constant with increasing pulse numbers. Also, the electroporated tissue area increased logarithmically with pulse number. Experimental and computational results on an IRE potato model were presented in [13] using the *standard EP model* with an experimental EP threshold of 184 V/cm. The authors reported a linear increase of the electroporated tissue area with voltage and a logarithmic increase with pulse number or pulse length.

The influence of pulse number and duration in IRE protocols was analyzed by [14] showing that an EP threshold decrease increased the tumor electroporated area. A numerical study of GET efficiency presented by [15] showed that a combination of short high voltage (HV) and long low voltage (LV) pulses was optimal and that the EP threshold decreased nonlinearly. The simulation of an in vivo EP-based protocol with a time-dependent numerical model presented by [16] succeeded in describing correctly all the pulse parameters. The electroporated tissue area had a logarithmic time increase. [17] presented a study of an IRE protocol for determining the EP threshold in prostate tissue *in vivo*. The authors reported an average EP threshold in the range of 500-550 V/cm using the *standard EP model* with EP thresholds from in vivo measurements. They concluded that the prostate cancer EP thresholds reported are only valid for protocols that follow identical isoelectric conditions (pulse length, active tip length, applied voltage).

In summary, it is widely accepted that EP threshold varies greatly with the type of tissue, time of exposition, experimental setup, and many other factors. In particular, many authors referenced in the bibliography reported experimental measurements of the EP threshold at the last pulse, in EP-based protocols, showing that it decreased in time as a function of the pulse number. However, no explanation was found for the nature of this variation nor the existence of a complex interaction between electroporated and EP threshold trajectories. These trajectories need to be analyzed.

In EP-based protocols, unwanted tissue damage may arise due to irreversible EP, pH effects, or thermal effects (depending on the protocol type). Damage due to irreversible EP and temperature effects was considered by many authors, for instance, in [18] simulated results show that for a given range of pulse parameters heating might be significant for GET protocols due to DNA damage. Damage due to pH was primarily studied in Electrolyte Ablation (EA), another non-thermal ablative method that uses a low long pulse on a tissue to generate electrolytic products inducing tumor necrosis [19]. Further studies of EA were presented among others in [20]. The results of pH damage in EA were extended to EP-based protocols in a series of works, among others in [21 and 22]. The concept of optimal dose response in a GET protocol in terms of pulse number was discussed in [23]. As previously discussed, when looking for an optimal EP-based protocol in terms of pulse number, the EP threshold trajectory is needed but it is seldom available, except at the last pulse. [23] proposed an exponential time decrease function to approximate the EP threshold trajectory, and [24] a methodology for approximating the exponential time decrease function.

Our goal is to introduce a new methodology, based on dynamical systems theory, to reveal the nonlinear dynamic interaction of the trajectories involved in an EP-based protocol, and to predict the threshold and electroporated area trajectories by measuring the electroporated area at the last pulse, thus avoiding the need to measure the EP threshold trajectory, a rather heavy task. The methodology is applied for the obtention of an optimal dose response in terms of pulse number in an EP-based protocol, considering damage due to pH, thermal, or irreversible EP effects.

The plan of the paper follows: Section 2 presents materials and methods; Section 3, the new methodology for predicting the electroporated tissue area trajectory in the context of an in vitro model (potato model); Section 4, its application for EP-based protocol optimization considering damage due to pH, thermal, or irreversible EP effects through an in vivo model (skinfold chamber); and Section 5, some general conclusions.

## Material and methods

An in vitro potato model and an in vivo dorsal skinfold model, in sections 3 and 4, respectively, are used to illustrate the genesis of the new methodology and its application on an EP-based protocol optimization. For conciseness, in what follows, by the area, we mean tissue area. Moreover, following the terminology used in nonlinear dynamical systems, by the electroporated area trajectory we mean electroporated tissue area variation in space and time, and by threshold trajectory we mean EP threshold variation in space and time. We remember that the electric field distribution as well as the threshold trajectory are defined by two elements, the set of their isoline closed curves and their value.

### The potato model (in vitro)

The in vitro model taken from [12] consists of the application of an EP-based protocol (32 pulses with 100 µs, 1 Hz, and 1500 V/cm) to a potato model with a six-electrode arrangement (Fig. 1, left), and the recording of the electric current and electroporated tissue area trajectory. After the application (24 hours), the potato exhibits an oxidation process revealed by darkened areas around the electrodes. Figure 1 right shows the darkened area after the last pulse. The oxidation is the product of the release of intracellular enzymes, thus implying an electroporated tissue area.

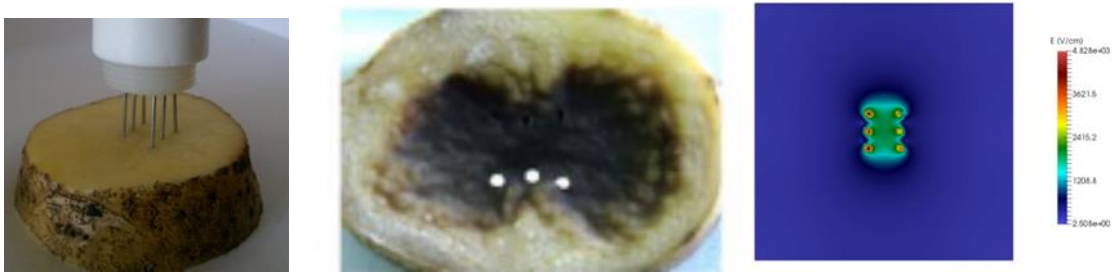

Figure 1. Experimental setup (left), the potato darkened area after the last pulse (center), simulated electric field distribution (right) for a pulse of 1500 V/cm. In the center figure, only one set of parallel electrodes is indicated by white dots. Left and center figures reproduced from [12].

### The potato model (in silico)

According to the *Standard EP model,* predicting the outcome of an EP-based protocol (in this case applied to the potato model) in terms of electrical variables consists of finding the extent of the electroporated tissue area that is, computing the electric field, and matching it with a known threshold. In mathematical terms, this translates into solving for each time step the nonlinear Poisson equation for the electric field distribution (with a tissue conductivity made a function of the electric field, temperature, and pulse number) and the time-dependent bioheat equation for the temperature distribution, and it's matching with the threshold from experimental measurements. Details of the equations and the electroporated tissue area prediction are found in the appendix, and in section 3, respectively.

The computational model solved the nonlinear Poisson equation and the time-dependent bioheat equation using finite elements. The domain was discretized with four-node elements in a 2D space and an ODE solver for the time domain. Several meshes with densities between 60000 and 180000 nodes were used. Natural boundary conditions on all external nodes were used for the Poisson equation, while Dirichlet boundary conditions (fixed to room temperature) were used for the bioheat equation. Figure 1 (right) shows an example of the results for the simulated electric field distributions (for a pulse of 1500 V/cm).

## The dorsal skinfold model (in vivo)

The dorsal skinfold model and pH measurements consist of the application of a GET protocol (12 pulses of 40 V/cm, 20 ms, and 1 Hz) to a transparent chamber implanted on a mouse dorsal skinfold with a two-needle configuration, and the recording of the electric current and the pH front propagation [21]. Intravital microscopy was performed with fluorescent microscopy and pH front visualization with phenol red as acid as well as basic pH indicators. Figure 2 shows a remarkable 2D snapshot of pH fronts, thus damage, induced by GET in mice through a dorsal skinfold.

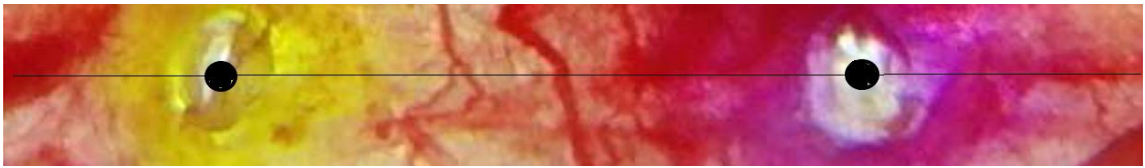

Figure 2: Intravital microscopy image of pH fronts induced by GET protocol (12 pulses of 40 V, 20 ms at 1Hz) in a dorsal skinfold chamber. pH front visualization with phenol red (acid front, yellow, and basic front, pink; red stripes are the capillary network; black circles indicate electrodes separated 2 mm. Image adapted from [21].

## The dorsal skinfold model (in silico)

The mathematical model used in the search for an optimal GET protocol in the dorsal skinfold chamber consists of two parts: the first one describes the optimal dose response without considering damage due to pH, and the second one considers it.

In the first part, the main goal consists of predicting the electroporated tissue area trajectory (the zone between electrodes) in a GET protocol (12 pulses of 200 V/cm, 20 ms at 1 Hz) applied to a skinfold chamber (using two needle electrodes separated 2 mm from one another). The model simulates the dorsal skinfold chamber as a 1D electric conductor and solves the nonlinear Poisson equation for the electric field distribution (with a tissue conductivity made a function of the electric field, temperature, and pulse number) and the time-dependent bioheat equation for the temperature distribution, and its matching with the threshold trajectory. The parameters for the skinfold model are taken from [23]. Details of the equations, and the electroporated tissue area trajectory prediction, are found in the appendix, and in Section 4, respectively. Once the electroporated tissue area trajectory is obtained, the protocol optimization is carried over as shown in Section 4. The nonlinear Poisson equation and the time-dependent bioheat equation were discretized with a tetrahedral finite element mesh in a 3D space and an ODE solver for the time domain. The computational model was implemented with COMSOL Multiphysics.

This second part addresses the optimal dose-response considering damage due to pH. pH fronts trajectory is obtained through the solution of the Nernst-Planck equations describing the electric field and ion transport in a seven-component species solution (H+, OH_, Cl_, Na+, HCO_3, CO2, and CO2_3). The skinfold chamber geometry configures a 3D problem; to reduce complexity, a 1D approximation was chosen in which the electric field trajectory varies over the segment joining the two electrode centers (see Fig. 3). The Nernst-Planck equations are described in [22]. The computational model approximates the skinfold chamber with a 1D grid containing the region between electrodes. The Nernst-Planck equations are approximated with finite differences and solved with standard relaxation techniques. The advancement of the pH fronts determines the induced damaged area (the area subjected to pH thresholds higher than 8.5 or lower than 4.5). Details of the computational results can be found in [22]. Having the damaged area trajectory, the protocol optimization follows (as shown in section 4).

# The new methodology for predicting the electroporated tissue area trajectory

In the following, we first recall the well-known classical method for predicting the threshold at the last pulse on an EP-based protocol by measuring the electroporated tissue area at the last pulse and using the *Standard EP Model*. Then we describe the new method in two stages, starting with an extension of the classical method that allows predicting the threshold trajectory, and ending with an extension of the latter that allows predicting the electroporated area trajectory with the sole measurement of the electroporated area at the last pulse.

In the classical method, the *Standard EP model* is routinely applied via the solution of the steady state nonlinear Laplace and the Bioheat equations (see appendix), for predicting the electroporated area at the last pulse when the threshold is known. Conversely, the *Standard EP Model* can be used to predict the threshold at the last pulse when the electroporated area is known. Here, the threshold is obtained via the matching of the electric field distribution and the known electroporated area. Note that the electric field distribution is used both ways, whether to predict the threshold knowing the electroporated area or vice versa.

Now, nothing impedes extending the classical method, seeing the EP-based protocol as a dynamical system that is a succession of steady states, and using the *standard EP model* via the solution of the nonlinear coupled system of the Laplace and the bioheat equations, for predicting the threshold trajectory provided the electroporated area trajectory is measured. This is the first stage of the new methodology.

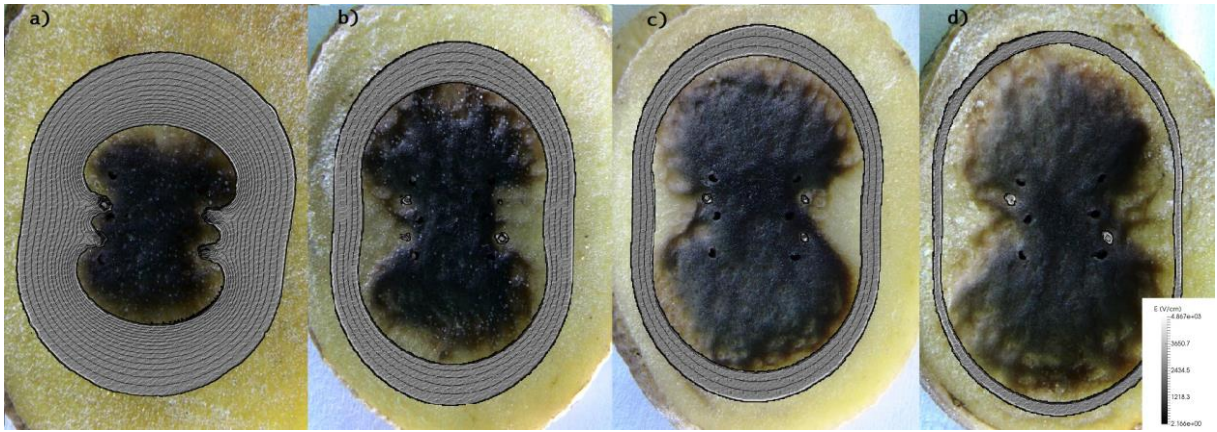

Figure 3: Darkened area trajectory, and the electric field distribution (isoline values below the threshold for EP). Innermost curves determined by the boundaries of the darkened areas configure the threshold isoline trajectory. The matching of the threshold isoline trajectory with the isolines of the electric field distribution at pulses a) 4, b) 8, c) 16, and d) 32 determines the value of the threshold isoline trajectory.

This is illustrated with the potato model taken from [12]. Figure 3 shows 4 snapshots of the darkened area trajectory at the 4th, 8th, 16th, and 32nd pulse numbers, obtained with a pulse intensity of 1500 V/cm and a frequency of 1 Hz. The darkened area trajectory appears to increase logarithmically in time. The problem here is finding the threshold trajectory, knowing the darkened area trajectory. Figure 3, depicts the measured darkened area trajectory and superposed in each pulse, the electric field distribution in the range 4-32 (only shown electric field isolines below the threshold for EP), coming from the solution of the nonlinear Laplace and Bioheat equations (see appendix for details). The time scale is fixed by the length of time during which the electric field values are above the threshold for electroporation, that is, 32 pulses. Note that the innermost curves in Fig. 3, determined by the boundaries of the darkened areas trajectory also configure the threshold isoline trajectory (though their threshold values are not yet known). But Fig. 3, also depicts the result of matching the threshold isoline trajectory with the isolines of the electric field distribution (at pulses 4, 8, 16, and 32), thus determining the value of the threshold isoline trajectory. This completes the threshold trajectory information. It is remarked that the set of thresholds at each pulse number in the range of 4-32 configures the threshold trajectory and that this threshold trajectory is the subset of the electric field having values above the critical electric field for EP. The threshold trajectory is called the electric field time gradient. The highest threshold value (and lowest isoline arc length) is at the first pulse number (close to the electrodes though not shown), while the lowest threshold value (and highest isoline arc length) is at the last pulse (shown as the innermost curves). The gray curve in Fig. 4, a time cross-section from the threshold values from Fig. 3 shows the threshold trajectory vs. pulse number for a cross-section parallel to the electrodes (later discussed). This curve is also the electric field time gradient. The threshold trajectory has an approximate exponential time decrease form. Note that the previous description uses well-known concepts in the electroporation field and new concepts from dynamical systems theory, such as electroporated area trajectory, electroporated area boundary isoline, threshold isoline trajectory, threshold trajectory, and electric field time gradient. As in dynamical systems, the study of trajectory interaction in electroporation opens new ways to understand the complex relation between electroporated area, electric field distribution, and threshold trajectories.

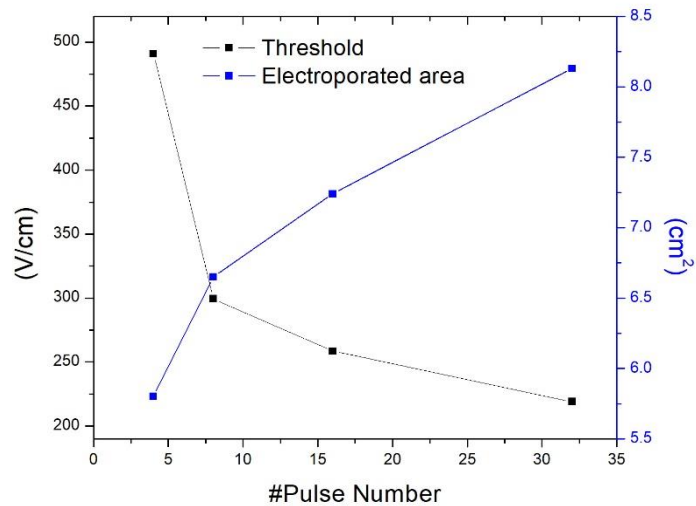

Figure 4: Threshold trajectory (gray) and darkened area
trajectory (blue) vs. pulse number

We are now ready to complete the description of the new methodology, that is, its second stage. The problem here is how to predict the darkened area trajectory in the potato EP-based protocol model, knowing the darkened area at the last pulse. As already seen, the electric field distribution for each pulse number in Fig. 3 is obtained from the numerical solution of the nonlinear Laplace and bioheat equations. The threshold isoline at the last pulse (pulse 32) is also known, thus fixing the time scale of the threshold isoline trajectory. However, the predicted threshold isoline trajectory is still unknown. Because the electric field is inhomogeneous, radial cross-sections of the electric field at its maximum value (in the range 4-32) yield a spectrum of electric field gradients. The steepest gradient is in the normal direction to the electrodes; the most gradual gradient is in the parallel direction to the electrodes. This gives upper and lower bounds of the electric field gradients. Here, we have chosen the lower bound corresponding to the parallel direction to become the electric field time gradient. In the context of EP-based protocol optimization, the chosen direction influences the protocol optimization. Having the threshold trajectory as the electric field time gradient, thus the isoline threshold trajectory, the electroporated area follows. In Fig. 4, the gray curve shows the threshold trajectory vs. pulse number for a cross-section parallel to the electrodes, and the blue curve, the electroporated area trajectory vs. pulse number. A comparison of the latter with the potato model experimental electroporated trajectory from [12] shows that the best option seems to be the parallel to the electrode's cross-section. Further experiments are needed to elucidate this point.

In synthesis, the salient features of the new methodology are the following. First, while the electric field remains constant after the first pulse, spanning from its highest value near the electrodes to zero at the boundary domain, the threshold trajectory (a subset of the electric field at each pulse) spans from its highest value at the first pulse to its lowest critical value at the last pulse. The threshold trajectory is the electric field time gradient. Thus, the threshold trajectory can be predicted from the electric field time gradient, knowing the electroporated area at the last pulse. In the case of inhomogeneous electric field space distribution, the threshold trajectory can be predicted from a spectrum of electric field time gradients: the steepest gradient is in the normal direction to the electrodes; the most gradual gradient is in the parallel direction to the electrodes. This gives upper and lower bounds of the electric field gradients and a choice is compulsory. Second, being the threshold at each pulse a subset of the electric field, the threshold trajectory must follow an approximate exponential time decrease. Third, since the darkened area trajectory's boundaries coincide with the threshold trajectory's isolines, they must increase in time in an approximate logarithmic form. A macroscopic explanation of the threshold trajectory behavior is that since the isoline of the electric field threshold (i.e. the boundary of the electroporated area) increases in an approximate logarithmic form, its value must decrease in an approximate exponential form. Another plausible reason is that the threshold trajectory, as a subset of the electric field, must follow an approximate inverse square law (which is close to an exponential time decrease). In summary, the new methodology provides a macroscopical explanation of why the threshold trajectory decreases as a function of the pulse number.

It is worth noticing the difference between the classical methodology extension that predicts the threshold trajectory knowing the electroporated measured area trajectory, and the new methodology that predicts the electroporated area trajectory knowing the measured electroporated area at the last pulse. In the former, the predicted threshold trajectory comes from experimental measurements of the electroporated area boundary isoline trajectory and its matching with the electric field isolines. In the latter, the predicted electroporated area trajectory comes from the electric field time gradient, there is no experimental input, except at the last pulse. Also, if the electric field is inhomogeneous the electric field gradient must be chosen beforehand. This implies some trial and error procedure.

## An optimal EP-based protocol in terms of pulse number

An optimal EP-based protocol in terms of pulse parameters, is a function of pulse amplitude, length, number, and frequency, among other variables. Finding an optimal protocol in such a large parameter space is a daunting task. [23] and [24] presented an optimal GET protocol for the dorsal skinfold model in terms of pulse number. In these works, the threshold trajectory was a priori assumed to follow an exponential time-decreasing function; unwanted damage due to pH fronts and the concept of the dose-response relationship was discussed. Here we look for an optimal GET protocol for the dorsal skinfold model in terms of pulse number by applying the new methodology introduced in Section 3. It consists of two parts: the first describes the optimal dose-response without considering damage due to pH, and the second considers it.

In the first part, a GET protocol (12 pulses of 200 V/cm, 20 ms at 1 Hz) is applied to a dorsal skinfold chamber (see Fig. 2). The computational model uses a 1D geometry defined along the zone between both electrodes, whose separation is 2 mm. The problem here is how to predict the electroporated area trajectory, knowing the electroporated area at the last pulse, i.e., assuming that the region between electrodes was completely electroporated after twelve pulses. The following assumptions are made with the sole purpose of illustrating the new methodology in a more simple setting. Due to the dorsal skinfold model two-electrode configuration, the electric field is inhomogeneous, here we assumed symmetry of the electric field distribution along a cross-section normal to the line connecting both electrodes at its midpoint. Moreover, we assumed that the electric field is radially homogenous. Following the methodology introduced in Section 3, Fig. 5 first row shows what it is known as a priori: the electric field distribution for each pulse number (gray curve) coming from the numerical solution of the nonlinear Laplace and bioheat equations, and the threshold isoline (violet line) at the last pulse (12). The latter fixes the time scale of the threshold isoline trajectory. However, the predicted threshold isoline trajectory is still unknown.

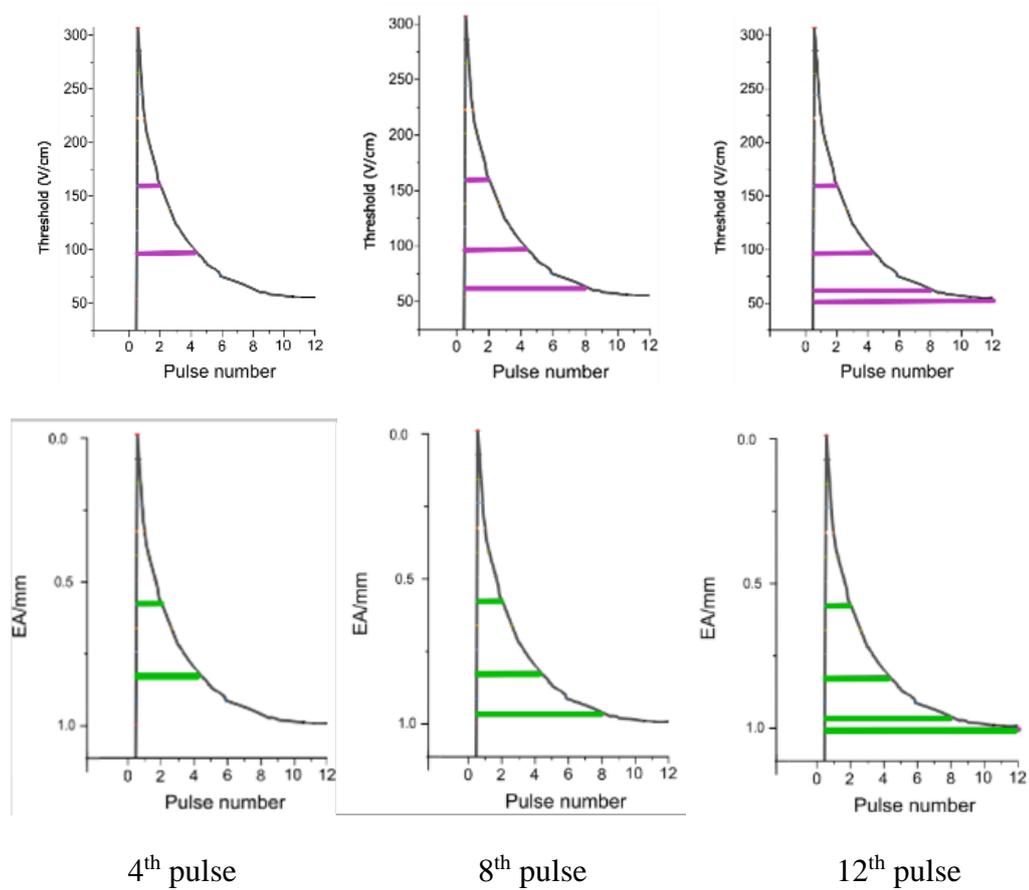

4th pulse          8th pulse          12th pulse

Figure 5: First row: Graph of the threshold isoline trajectory (gray curve), and electric field isolines (violet lines), for pulse numbers 4, 8, and 12. Second row: Graph of the electroporated boundary area trajectory (gray curve), and electroporated area trajectory (green lines), for pulse numbers 4, 8, and 12.
Note the inverted ordinate axis in Fig. 5 second row.

The radial cross-section of the homogeneous electric field distribution yields equal electric field gradients. Therefore, any electric field radial cross-section becomes the electric field time gradient, thus the threshold isoline trajectory and its values (the whole threshold trajectory is seen at the last pulse and it is constituted by the gray curve and the set of violet isolines). Having the threshold trajectory as the electric field time gradient, thus the isoline threshold trajectory, the electroporated area follows. Figure 5 second row, shows a graph of the electroporated area trajectory (green lines) increasing in time. Note that the graphs are upside down (the origin is at the top where the threshold is the maximum) to show that, at the same pulse number, the boundary of the electroporated tissue area and the threshold isoline are identical geometrical objects. The green curve in Fig. 6 depicts the predicted trajectory vs. pulse number with the ordinate reversed. Again, the predicted threshold trajectory follows an approximate exponential time decrease, while the electroporated area follows an approximate logarithmic time increase. Note that this is the result of assuming that the region between electrodes was electroporated after twelve pulses, which in this example is approximately 50 V/cm. Had we chosen a threshold minimum value of 100 V/cm after twelve pulses (which according to the literature, is more realistic), the threshold trajectory would comprise the electric field values between the maximum value (approximately 300 V/cm) and the minimum 100 V/cm (in this case, more than 70% of the region between electrodes will not be electroporated).

As in Section 3, the main conclusion is that it is possible to predict the threshold isoline trajectory and the electroporated area trajectory with the measurement of a single value of the electroporated area. The threshold isoline and the electroporated boundary area trajectories are identical geometrical objects. Further, the threshold value trajectory has an approximate exponential time decay function while the threshold isoline trajectory, i.e. the electroporated area boundary trajectory, has an approximate logarithmic time increase function.

Having the electroporated area trajectory as a function of the pulse number (as shown in Fig. 6), the optimal pulse dosage (pulse number), is defined as the first pulse number in which the maximum difference between two successive values of the electroporated tissue area trajectory [TT (n+1) − TT(n)] is smaller than a prescribed small value [23] (here TT stands for treated tissue area). It is optimal because the increase in pulse number is not followed by an increase in the electroporated tissue area. In this example, the optimal pulse number would be around 12.

This definition deals with the optimal dose of the protocol itself, independent of the existence of a target. However, if a target exists such as in an ECT protocol, the optimal dose would be the first pulse dosage in which the maximum difference between the electroporated tissue area and the target tissue area is smaller than a prescribed small value.

Now we couple the new methodology from Section 3 using the *standard EP model* with the *standard EA model* to illustrate the interaction of electroporated and damaged area trajectories in the prediction of an optimal dose-response relationship in a GET protocol considering tissue damage due to pH. In what follows, the term *treated area* trajectory is analog to the *darkened area trajectory* (in the context of the potato model), *damaged area trajectory* is the area around electrodes under critical pH values, and the *reversibly electroporated area trajectory* is the difference between *treated area* and *damaged area* trajectories. In the coupling of the *standard EP model* and the *standard EA model*, the former has been previously described, and the latter assumes tissue as a 1D ionic conductor and solves the nonlinear Nernst-Planck equations governing the ion transport between electrodes, see details in [22]. The pH space-time distribution defines the pH-damaged area trajectory. This is discussed in [22] which shows the predicted pH-damaged trajectory from a GET protocol (12 pulses of 200 V/cm, 20 ms, and 1 Hz). Also, when the damaged area is considered, the dose becomes the pulse dosage, and the response becomes the reversibly electroporated and unwanted pH-damaged area trajectories. Figure 6 illustrates how the optimal dose response is obtained through the coupling of both models. The pH-damaged area trajectory represented by the red curve is taken from [22]).

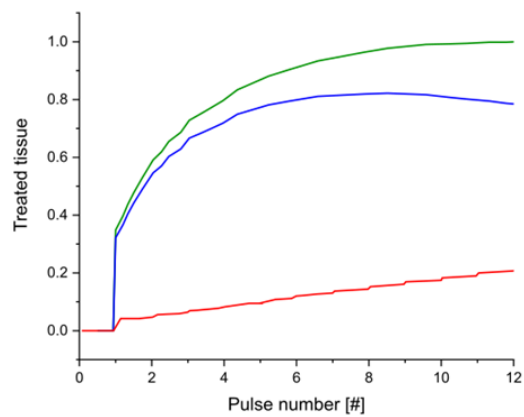

Figure 6: The predicted reversible electroporated area trajectory (blue) is the treated area trajectory (green) minus the pH-damaged area trajectory (red).

In Fig. 6, the green curve is the former electroporated area trajectory from Fig. 5 (with the scale inverted), now called the treated area trajectory. The blue curve becomes the electroporated area trajectory, that is, the difference between the treated area trajectory (green) and the pH-damaged trajectory (red). The green curve scales approximately logarithmically and the red curve linearly, their difference yields a maximum in the blue curve. The pulse number at this maximum is the optimal pulse dosage, past this point, the reversibly electroporated area trajectory decreases while the damaged area trajectory increases. In this example, the optimal pulse number would be around 8. Comparing the results of the GET protocol with and without pH effects shows that the damage reduces the optimal pulse dosage, but also the reversibly electroporated area. It is noted that the damaged area trajectory due to temperature effects varies similarly to the pH-damaged area trajectory, thus the first part of our methodology can be extended to the optimization of GET protocols, considering temperature area damage. This is also valid for the optimization of IRE protocols, provided the dose-response is correctly defined. If the IRE effect consists of apoptotic cell death, the dose is the pulse dosage, and the response is the irreversibly electroporated area trajectory and the unwanted damaged area trajectory due to temperature or pH effects.

## Conclusions

In the search for optimal dose response in terms of pulse number in EP-based protocol optimization in terms of pulse number, it is crucial to predict the electroporated and damaged area trajectories. This requires measuring the threshold trajectory and analyzing the interaction of electroporated, threshold and damaged area trajectories. Here we introduced a new methodology, based on the theory of nonlinear dynamical systems interaction, that shows that the threshold trajectory is the time gradient of the electric field. This makes it possible to predict the electroporated area trajectory by measuring the electroporated area at the last pulse, thus avoiding the need to measure the EP threshold trajectory, a rather heavy task. Some of the salient features of the new methodology are: in the case of an EP-based protocol having a homogeneous electric field space distribution, the threshold trajectory can be predicted from the time gradient of the electric field time; if the electric field distribution is inhomogeneous, the threshold trajectory can be predicted from a spectrum of electric field time gradients with an upper and lower bound of the electric field gradient, a choice is compulsory; the threshold trajectory aka the electric field time gradient, is a subset of the electric field gradient, thus explaining its approximate exponential time decrease: also, the electroporated boundary area trajectory, i.e. the threshold isoline trajectory, follows an approximate logarithmic time increase; this provides a macroscopical explanation of why the threshold trajectory decreases as a function of the pulse number; another plausible reason is that the threshold trajectory, as a subset of the electric field, must follow an approximate inverse square law (which is close to an exponential time decrease). In summary, the new methodology provides a macroscopical explanation of why the threshold trajectory decreases as a function of the pulse number.

It is worth noticing the difference between the extension of the classical methodology that predicts the threshold trajectory knowing the electroporated area trajectory, and the new methodology that predicts the electroporated area trajectory by measuring the electroporated area at the last pulse. In the former, the predicted threshold trajectory comes from experimental measurements of the electroporated area boundary isoline trajectory and its matching with the electric field isolines. In the latter, the predicted electroporated area trajectory comes from the electric field time gradient, there is no experimental input, except at the last pulse. Also, if the electric field is inhomogeneous the electric field gradient must be chosen beforehand. This implies some trial and error procedure.

A comparison between the 1D approximation of the dorsal skinfold and the 2D potato model shows that in the former, the 1D approximation and the symmetry assumption result in a radially uniform electric field gradient, thus the electric field time gradient and the threshold trajectory. In the latter, the electric field trajectory is non-uniform, and the threshold trajectory obtained has a spectrum of values between a maximum and a minimum electric field time gradient. The choice of the electric field gradient has a direct consequence on the predicted electroporated area trajectory and in the obtention of an optimal pulse dosage in EP-based protocols.

When area damage is considered, a third component is included in the interaction analysis: the damaged area trajectory. As shown in the results, the interaction of the damaged and electroporated area trajectories has definite effects on the latter, and consequently, on the EP-based protocol optimization. Moreover, comparing the electroporated area trajectory from the new methodology with those from [23] that approximates the threshold trajectory with an exponentially time-decreasing function, shows a steeper threshold isoline trajectory in the first pulses in the former. Since in the new methodology, the threshold isoline trajectory is the electric field time gradient (thus coming from first principles), rather than an a priori heuristic approximation, the results presented here should be more reliable. An upshot of the methodology introduced is that many controversial EP threshold values from the literature are reconciled by figuring out the closeness of their threshold trajectories, which implies the closeness of the experimental setup and electric parameters. It is hoped that the methodology introduced would help in designing optimal treatment planning in electrochemotherapy (ECT), irreversible electroporation (IRE), and gene electrotransfer (GET).

## Acknowledgments

G. Marshall is Senior Researcher at the Consejo Nacional de Investigaciones Científicas y Técnicas (CONICET). This work was supported by grants from CONICET PIP 2017/22, and Universidad de Buenos Aires UBACYT 2017/22, ANPCYT 2017/22. The founders had no role in the study, design, data collection, analysis, the decision to publish, or preparation of the manuscript. G. M. thanks Isaac Rodriguez Osuna for his help with the COMSOL computer modeling, and Cecilia Suárez for providing experimental data from the potato model.

## References

1. D. Miklavcic, Handbook of Electroporation, Springer International Publishing, Cham, 2017, https://doi.org/10.1007/978-3-319-32886-7. http://link.springer.com/10.1007/978-3-319-32886-7.

2. E. Neumann, M. Schaefer-Ridder, Y. Wang, P.H. Hofschneider, Gene transfer into mouse glioma cells by electroporation in high electric fields, EMBO J. 1 (7) (1982) 841-845, https://doi.org/10.1385/1-59259-409-3:55.

3. L.M. Mir, M.F. Bureau, J. Gehl, R. Rangara, D. Rouy, J.M. Caillaud, P. Delaere, D. Branellec, B. Schwartz, D. Scherman, High-efficiency gene transfer into skeletal muscle mediated by electric pulses, Proc. Natl. Acad. Sci. U.S.A. 96 (8) (1999) 4262-4267, https://doi.org/10.1073/pnas.96.8.4262

4. J. Gehl, L. Mir, Determination of optimal parameters for in vivo gene transfer by electroporation, using a rapid in vivo test for cell permeabilization, Biochem. Biophys. Res. Commun. 380 (1999).

5.  Gehl, J., Sørensen, T. H., Nielsen, K., Raskmark, P., Nielsen, S. L., Skovsgaard, T., and Mir, L. M. (1999)] In vivo electroporation of skeletal muscle: threshold, efficacy, and relation to electric field distribution Biochim. Biophys. Acta 1428 (1999).

6.  D. Miklavcic, D. Semrov, H. Mekid, and L. M. Mir, "A validated model of *in vivo* electric field distribution in tissue for electrochemotherapy and for DNA electrotransfer for gene therapy," *Biochim. Biophys. Acta*, V 1523 (2000).

7.  Davorka Sel, David Cukjati, Danute Batiuskaite, Tomaz Slivnik, Lluis M. Mir, and Damijan Miklavcic, Sequential Finite Element Model of Tissue Electropermeabilization IEEE Transactions on Biomedical Engineering, vol. 52, No. 5, 2005 DOI: 10.1109/TBME.2005.845212.

8.  Jiang C, Davalos RV, Bischof JC (2015) A review of basic to clinical studies of irreversible electroporation therapy. IEEE Trans Bio-Med Eng 62(1):4–20.

9.  G. Pucihar, J. Krmelj, M. Reberscek, T. Napotnik, D. Miklavcicc, Equivalent pulse parameters for electroporation, IEEE (Inst. Electr. Electron. Eng.) Trans. Biomed. Eng. 58 (11) (2011), https://doi.org/10.1109/tbme.2011.2167232.

10. C.B. Arena, C.S. Szot, P.A. Garcia, M.N. Rylander, R.V. Davalos, A three-dimensional in vitro tumor platform for modeling therapeutic irreversible electroporation, Biophys. J. 103 (9) (2012) 2033-2042, https://doi.org/10.1016/j.bpj.2012.09.017.

11. S. Čorović, A. Županič, and D. Miklavčič, 'Numerical Modeling and Optimization of Electric Field Distribution in Subcutaneous Tumor Treated With Electrochemotherapy Using Needle Electrodes',  doi: 10.1109/TPS.2008.2000996.

12. C. Suarez, A. Soba, F. Maglietti, N. Olaiz, G. Marshall, The role of additional pulses in electropermeabilization protocols, PLoS One 9 (12) (2014), e113413, https://doi.org/10.1371/journal.pone.0113413.

13. Q. Castellví, J. Banús, A. Ivorra, 3D assessment of irreversible electroporation treatments in vegetal models, IFMBE Proc. 53 (2016) 294-297, https://doi.org/10.1007/978-981-287-817-5_65

14. P.A. Garcia, B. Kos, J.H. Rossmeisl Jr., D. Pavliha, D. Miklavcic, R.V. Davalos, Predictive therapeutic planning for irreversible electroporation treatment of spontaneous malignant glioma, Med. Phys. 44 (9) (2017) 4968e4980.

15. T. Forjani, D. Miklavcic, Numerical study of gene electrotransfer efficiency based on electroporation volume and electrophoretic movement of plasmid DNA, Biomed. Eng. Online 17 (2018) 80, https://doi.org/10.1186/s12938-018-0515-3

16. M. Pintar, J. Langus, I. Edhemovic, E. Brecelj, M. Kranjc, G. Sersa, T. Sustar, T. Rodic, D. Miklavcic, T. Kotnik, B. Kos, Time-dependent finite element analysis of in vivo electrochemotherapy treatment, Technol. Canc. Res. Treat. 17 (2018 Jan 1), https://doi.org/10.1177/1533033818790510

17. Matthijs J. Scheltema M. J et al.   Numerical simulation modeling of the irreversible electroporation treatment zone for focal therapy of prostate cancer, correlation with whole-mount pathology and T2-weighted MRI sequences *Ther Adv Urol* 2019, Vol. 11: 1–10 DOI: 10.1177/1756287219852305

18. I. Lackovic, R. Magjarevic, D. Miklavcic, Three-dimensional finite-element analysis of joule heating in electrochemotherapy and in vivo gene electrotransfer, IEEE Trans. Dielectr. Electr. Insul. 16 (5) (2009) 1338e1347, https://doi.org/10.1109/TDEI.2009.5293947

19. E. Nilsson, H. Von Euler, J. Berendson, A. Thorne, P. Wersall, I. Naslund, A.S. Lagerstedt, K. Narfstrom, J.M. Olsson, Electrochemical treatment of tumors, Bioelectrochem. Bioenerg. 51 (1) (2000) 1-11, https://doi.org/10.1016/S0302-4598(99)00073-2

20. L. Colombo, G. González, F. Marshall, A. Soba, C. Suárez, P. Turjanski, Bioelectrochemistry 71 (2007) 223.

21. N. Olaiz, E. Signori, F. Maglietti, A. Soba, C. Suarez, P. Turjanski, S. Michinski, G. Marshall, Tissue damage modeling in gene electrotransfer: the role of pH, Bioelectrochemistry 100 (2014) 105-111, https://doi.org/10.1016-j.bioelechem.2014.05.001.

22. M. Marino, N. Olaiz, E. Signori, F. Maglietti, C. Suarez, S. Michinski, G. Marshall, pH fronts and tissue natural buffer interaction in gene electrotransfer protocols, Electrochem. Acta 255 (Supplement C) (2017) 463-471 https://doi.org/10.1016/j.electacta.2017.09.021

23. Luján, E., Marino, M., Olaiz, N. & Marshall, G. Towards an optimal dose-response relationship in gene electrotransfer protocols. Electrochem. Acta **319**, 1002–1011. https ://doi.org/10.1016/j.elect acta.2019.07.029 (2019).

24. M. Marino, E. Luján, E. Mocskos, G. Marshall, OpenEP: an open-source simulator for electroporation-based tumor treatments, Scientific Reports (2021) 11:1423

# Appendix I

Concerning the potato model, according to the *Standard EP model,* predicting the outcome of an EP-based protocol in terms of electrical variables consists of finding the electroporated tissue area trajectory through the solution at each time step of the nonlinear Poisson equation for the electric field distribution (with a tissue conductivity made a function of the electric field, temperature, and pulse number) and the time-dependent bioheat equation for the temperature distribution, and its matching with the threshold trajectory from experimental measurements. The nonlinear Poisson equation reads:

$$\nabla(\sigma(E,T)\nabla\Phi) = 0 \qquad (1)$$

Where $\vec{E} = -\nabla\Phi$, φ is the electrostatic potential, E is the electric field, T is the temperature and $\sigma(E, T, p)$ is the mean electrical conductivity (defined below).

The bioheat equation reads:

$$\nabla(k\nabla T) + q^m + \sigma(E,T,p)|\nabla\Phi|^2 = \rho C_p \frac{\partial T}{\partial t} \qquad (2)$$

Here $q^m$ is the metabolic heat generation, ρ the tissue density, Cp is the tissue heat capacity, and t is the time. Here, the tissue electric conductivity is made a function of the electric field (E), the temperature (T), and the number of pulses (p).

$$\sigma(E,T,p) = \sigma_0\big[1 + F_f(E) + F_T(T) + F_t(U,p)\big] \qquad (3)$$

$\sigma_0$ is it's the mean basal electrical conductivity.

$$F_f(E) = \begin{cases} 3.5 E > E_{irrev} \\ 1.0 + 2.5 \, \frac{E-E_{rev}}{E_{irrev}-E_{rev}} E_{irrev} \geq E \geq E_{rev} \\ 1.0 E_{rev} > E \end{cases} \qquad (4)$$

$E_{irrev}$, and $E_{rev}$ are defined in [12], and

$$F_T(T) = a_T[T - a_T] \tag{5}$$

being $a_T$ the temperature coefficient, and $a_T$ the physiological temperature.

$$F_t(U,p) = a + bU + cp + dU^2 + ep^2 + fUp \tag{6}$$

The first term of the last equation is fitted to experimental measurements from the potato tissue, where a = 8.183, b = -0.013, c = 0.0286, d = $4.472.10^{-6}$, e = $1.2.10^{-8}$, and f = $8.413.10^{-5}$. More details, as well as the physical parameters used in the simulations, are given for the potato model in [12].

Concerning the skinfold model, predicting the outcome of an EP-based protocol in terms of electrical variables, according to the *Standard EA model,* consists of finding the damaged tissue trajectory through the solution at each time step of the Nernst-Plank equations governing ion transport. Details of the model, the equations, its numerical solution, and the physical parameters used are given in [21] and [23].